\documentclass[reprint,
tightenlines,
superscriptaddress,
 amsmath,amssymb,aps,pra
]{revtex4-2}
\usepackage{graphicx}
\usepackage{dcolumn}
\usepackage{bm}
\usepackage{natbib}
\usepackage{braket}
\usepackage{amsmath}
\usepackage{hyperref}
\hypersetup{colorlinks=true,citecolor=blue}
\usepackage{color,xcolor,soul}

\begin{document}

\title{Internal diffraction dynamics of trilobite molecules}

\author{Rohan Srikumar}
\email{rsrikuma@physnet.uni-hamburg.de}
\affiliation{Zentrum für Optische Quantentechnologien, Universität Hamburg, Luruper Chaussee 149, 22761 Hamburg, Germany}

\author{Seth T. Rittenhouse}
\email{rittenho@usna.edu}
\affiliation{Department of Physics, United States Naval Academy
, Annapolis, MD 21402 USA}
\affiliation{Institute of Theoretical Physics, Institute of Physics, University of Amsterdam, Science Park 904, 1098 XH Amsterdam, The Netherlands}

\author{Peter Schmelcher}
 \email{pschmelc@physnet.uni-hamburg.de}
\affiliation{Zentrum für Optische Quantentechnologien, Universität Hamburg, Luruper Chaussee 149, 22761 Hamburg, Germany}
\affiliation{The Hamburg Centre for Ultrafast Imaging, Universität Hamburg, Luruper Chaussee 149, 22761 Hamburg, Germany}

\begin{abstract}

Trilobite molecules are ultralong-range Rydberg molecules formed when a high angular momentum Rydberg electron scatters off of a ground-state atom. Their unique electronic structure and highly oscillatory potential energy curves support a rich variety of dynamical effects yet to be explored. We analyze the vibrational motion of these molecules using a framework of adiabatic wavepacket propagation dynamics and observe that for appropriate initial states, the trilobite potential acts as molecular diffraction grating. The quantum dynamic effects observed are  explained using a Fourier analysis of the scattering potential and the associated scattered wavepacket. Furthermore, vibrational ground-states of the low angular momentum ultralong-range Rydberg molecules are found to be particularly suitable to prepare the relevant wavepackets. Hence, we propose a time resolved pump-probe scheme designed for the realization of the effect in question, and advertise the utilization of a single diatomic Rydberg molecule as a testbed for the study of exaggerated quantum dynamical phenomena.

\end{abstract}

\maketitle

\section{Introduction} \label{Sec1}

Ultracold atoms in highly excited Rydberg states have proven to be an outstanding tool in the engineering and control of long-range interactions in atomic, molecular and optical systems \cite{Browaeys2010,Zeppenfeld_2017,Busche2017,Rittenhouse2010,Vaneeclo2022}. They offer several exploitable features including a highly tunable electronic character which is readily usable for mediating a plethora of atom-atom interaction effects \cite{Urban2009,Weber_2017,Matthias2010,Browaeys2013}. In addition, interactions between a Rydberg atom and a neutral ground-state atom \cite{Omont_1977} mediated by the Rydberg electron has seen a surge in popularity in the context of engineering and detection of bound molecular states \cite{Greene2000,Fey2019}, ultracold chemical reactions \cite{Pfau2016}, and measurement of spatial correlations \cite{Whalen_2019,Killian2023}.

Notably, the low-energy scattering of a Rydberg electron off of a ground-state atom can result in the formation of ultralong-range Rydberg molecules (ULRMs), which are bound scattering states that feature massive internuclear proportions \cite{Greene2000,Bendkowsky2009,Hummel2020}. Three main classes of ULRMs are low-$l$, trilobite, and butterfly molecules \cite{Hamilton_2002,Niederprum2016} classified by the low or high angular momentum $l$ state of the Rydberg electron. In contrast to energetically isolated, quantum defect splitted low-$l$ ULRM, trilobite molecules are formed when the nearly degenerate set of high-$l$ hydrogenic states of a Rydberg electron are hybridized by a scattering neutral atom called perturber. Butterfly molecules on the other hand, are formed as a consequence of $p$-wave shape resonances that exist in the low-energy scattering of an electron with an alkali metal atom \cite{Omont_1977}. These intriguing systems have undergone numerous theoretical and experimental investigations that explored their electronic and rovibrational spectra \cite{Shaffer2018,Killian2022,Thomas_2018,Bellos_2013,Sadeghpour_2011,Raithel2014,Kleinbach_2017,Raithel_2016,Niederprum2016,Butscher_2011,Deiss_2021,Althoen2023}, their responses to external fields \cite{Lesanovsky_2006,Kurz_2014,Hummel_2019,Keiler_2021,Kurz_2013}, nonadiabatic effects \cite{Hummel_2022,Srikumar2023,durst2024}, and the complex angular momentum structure they feature \cite{hummel2018,Eiles_2017,Hummel2020,Anderson2014,Ott2016,eiles2024}. Recently there has emerged an interest in the dynamics of the consituent nuclei under the influence of these potentials \cite{Keiler_2021,Fey2016}. Apart from the neutral trilobite molecule investigated here, nuclear wavepacket dynamics of charged Rydberg molecules have also garnered a great deal of recent interest \cite{Raithel2022,anasuri_2023}.

From the double slit experiment to Bragg scattering of electron from crystal lattices, diffraction and interference effects exhibited by quantum matter have been a matter of significant experimental and theoretical interest from the very inception of quantum mechanics.  More recent characterization of similar phenomena includes, among many others, quantum diffraction of confined matter-waves and localized wavepackets by time independent and time dependent external potentials \cite{Brukner1997,Arseni2012,Kälbermann_2001,Moshinky1952,Zecca2011}. Here, we propose a system where the electronic potential energy structure of a trilobite molecule mediates the scattering and diffraction of a vibrational wavepacket creating a new approach to observe quantum interference effects in a single diatomic molecular system of huge size. 

In this article, we demonstrate that the adiabatic potential energy curves (PEC) of the trilobite molecules form a unique diatomic environment to probe quantum dynamic effects in time and length scales not customarily accessible to conventional molecules. Apart from their exaggerated spatial extent and slow temporal evolution, the trilobite molecules also feature very high polarizability, extreme sensitivity to electric fields and a rapidly varying electronic character not reproducible in the dynamics of standard diatomic molecules. A detailed characterization of wavepacket dynamics of a ULRM can provide insight not just into interesting bound vibrational effects, but also low-energy $l$-changing collision between the ground-state atom and the Rydberg atom \cite{Hummel2021,Pfau2016}. The ULRM hence presents itself as a diatomic laboratory capable of performing yet to be explored cold-atom scattering experiments.
\begin{figure}
    \includegraphics[width=0.5\textwidth]{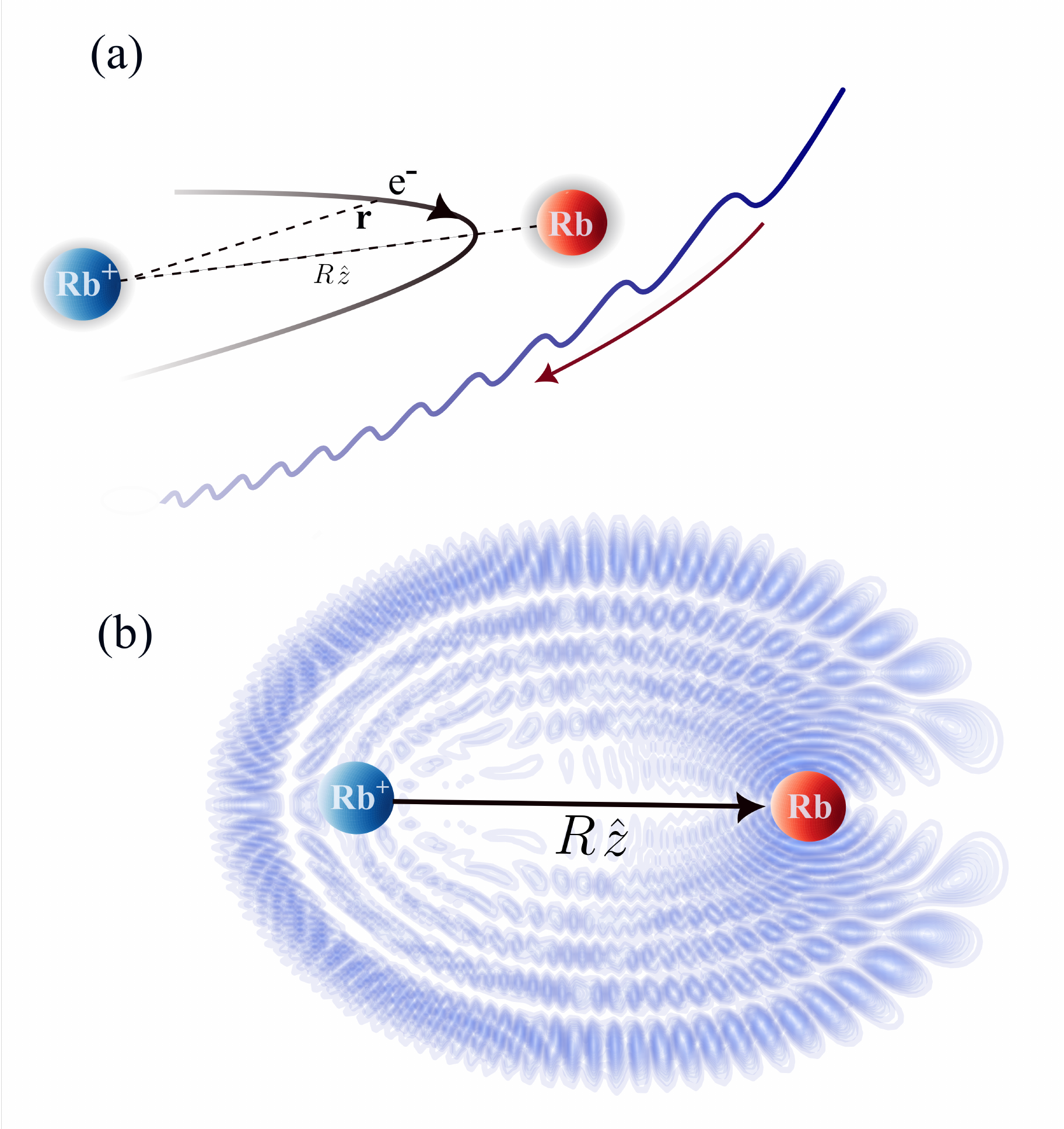}
    \caption{Illustrative sketch of a trilobite molecule. (a) The binding interaction between the two $^{87}$Rb atoms separated by an internuclear vector $R \hat{z}$ (along the z-axis) is determined by the low-energy scattering of the excited electron (positioned at $\mathbf{r}$) with the ground-state atom. (b) Characteristic electronic density of a $n=55$ trilobite state with $R=4000$ a.u.}
    \label{fig1}
\end{figure}

In this work we perform numerical simulations of the vibrational wavepacket dynamics of a trilobite molecule as it propagates through the associated trilobite electronic potential. We choose the diatomic $^{87}$Rb ULRM as our molecular prototype. We observe that for a range of values of the principal quantum number $n$, the adiabatic electronic potentials of the ULRM exhibit an undulating structure along the internuclear axis. These potential energy curves can act as a diffraction grating and are back-scattering appropriate incident wavepackets to form time propagating multi-peak structures.  To identify the underlying mechanism, we use a Fourier analysis of the PEC and analyze the behavior of the scattered wavepackets in both position and momentum space. We also show that the necessary initial state of the system can be conveniently prepared by the use of advanced optical tweezer setups, or alternatively initialised utilizing the vibrational ground-state of low-$l$ ULRM states.

This article is organized as follows. In section \ref{Sec2} we provide our theoretical framework describing the internal electronic structure of the ULRM (\ref{Sec21}), the potential energy landscape as well as the relevant nuclear dynamics (\ref{Sec22}). In Section \ref{Sec3} we present our numerical results,  with subsection \ref{Sec31} highlighting the energetical structure of the trilobite molecule, subsection \ref{Sec32} detailing its vibrational dynamics, \ref{Sec33} addressing potential limitations in the theoretical framework, and \ref{Sec34} detailing the elementary experimental schemes that could be the basis to observe the aforementioned vibrational dynamics. Lastly, our conclusions and future research outlook are provided in Section \ref{Sec4}.

\section{Theoretical framework and Methodology}    \label{Sec2}

We begin by characterizing the electronic interaction between the constituents of our system and model the electronic Hamiltonian for the ULRM. The salient features of the trilobite state are highlighted and their effects on the vibrational dynamics of the ULRM are discussed. We then present the time-dependent Schrödinger equation that governs the internal nuclear dynamics of the ULRM. Throughout our work, the system and the results are presented and analysed in atomic units (a.u.) unless explicitly mentioned otherwise.

\begin{figure*}
    \centering
    \includegraphics[width=\textwidth]{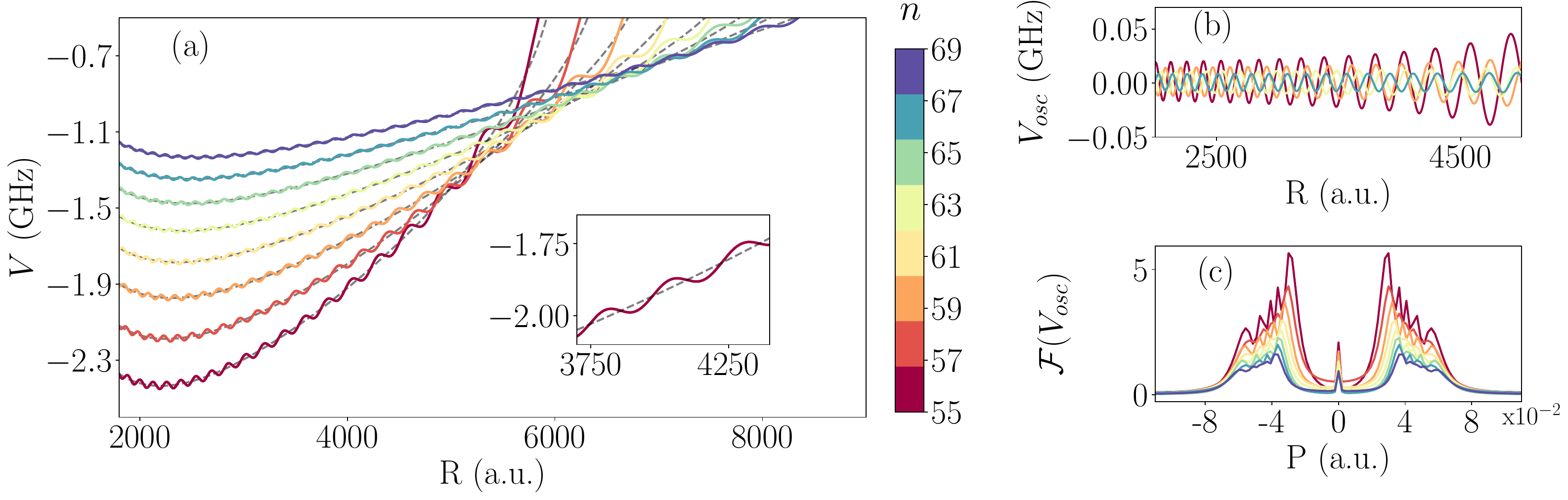}
    \caption{ Potential energy structure of trilobite molecule for ranges of $55 \leq n \leq 69$ color-coded with respect to $n$. (a) The colored solid lines portray the calculated trilobite potentials $V_t$ for multiple $n$-values, whereas the dotted grey lines plotted in the background represent the approximate curve envelopes $V_{anh}$ (inset magnified to highlight the structure of the PEC). (b) Represents the oscillatory feature of the trilobite potential, defined as $V_{osc}$, with (c) being the corresponding Fourier transforms $\mathcal{F}\{ V_{osc} \}$.  }
    \label{fig2}
\end{figure*}
\subsection{Electronic interaction and structure}    \label{Sec21}
We consider a system consisting of two $^{87}$Rb atoms, one of which is excited to a high-lying Rydberg state, bound to each other via the low-energy scattering of the corresponding Rydberg electron with the ground-state atom. The triplet s-wave scattering of the Rydberg electron off of the ground-state atom facilitates an attractive interaction between the two Rb atoms, which is effectively modeled using a Fermi pseudo potential \cite{Fermi1934,Omont_1977} given by:

\begin{equation}    \label{Eq:V_en}
    V_{en}(\textbf{r},\textbf{R}) = 2\pi a_s^T(k) \delta(\textbf{r}-\textbf{R}),
\end{equation}
where $\textbf{r}$ and $\textbf{R}$ are respectively the positions of the Rydberg electron and the ground-state atom relative to the positive Rydberg core (as seen in Fig.~\ref{fig1}(a)).  Here, $a_s^T(k) = a_s^T(0) + \pi \alpha k/3$, is the low-energy triplet s-wave scattering length for electron-atom collisions, with the zero-energy scattering length $a_s^T(0) = -$16.05 \cite{Greene2000,Bahrim2000}, the ground-state atom polarizability $\alpha$ = 319.2 \cite{molof1974}, and the excited electron's wave-number $k$, given semi-classically in atomic units as $k^2 = 2/R - 1/n^2$. The singlet scattering interaction, being weaker and repulsive is safely ignored \cite{Greene2000,Bahrim2000}.
The adiabatic electronic Hamiltonian for the ULRM is then given by $H_{\mathrm{el}} = H_{\mathrm{Ryd}} + V_{\mathrm{en}}$, with $H_{\mathrm{Ryd}}$ describing the Rydberg electron in the potential of its parent ionic core.

For the purpose of this work we extract the Born-Oppenheimer potentials for the ULRMs in the molecular body-fixed frame at the level of perturbation theory. A complete, non-perturbative treatment is possible \cite{Eiles_2017,eiles2024}, but only creates small quantitative changes and no qualitative change for the results presented here. We express the system in the basis formed by the Rydberg atomic eigenstates $\{ \phi_{nlm} (\textbf{r}) \}$ of Rubidium with energies given in atomic units by $E_{nl}=-0.5(n-\mu_l)^{-2}$ where $\mu_l$ is the quantum defect \cite{Gallagher2003}. In the body-fixed molecular frame, there is a node along the internuclear axis for all $m\ne0$ states which therefore do not experience the contact electron-atom interaction of Eq.~\eqref{Eq:V_en}. Rydberg states with lower angular momenta $l < l_{min}$ ($l_{min} =$ 3 for Rb), are energetically separated due to non-negligible quantum-defects. At the level of first-order perturbation theory in $V_{en}$ the related PECs are given by
\begin{equation} \label{Eq:V_l}
    V_l(R) = 2\pi a_{s}^T(k)|\phi_{n^* l}(R)|^{2},
\end{equation}
where $n^* = n - \mu_l$ is the effective principle quantum number.

The high-$l$ ($l\ge3$) electronic sates create an approximately degenerate manifold of hydrogenic Rydberg states and thus must be treated using degenerate perturbation theory. The resulting PEC are shown for a range of principal quantum numbers in Fig.~\ref{fig2}(a) and are given as \cite{Greene2000}
\begin{equation}    \label{Eq:V_t}
    V_t(R) = 2 \pi a_{s}^T(k) \hspace{1mm} \sum_{l=3}^{n-1} |\phi_{nl}(R)|^{2}.
\end{equation}
 The corresponding electronic state 
\begin{equation}    \label{Eq:trilobite}
    \psi_t(R,\mathbf{r}) = \frac{1}{T}\sum_{l=3}^{n-1}  \phi_{nl}(R) \phi_{nl}(\mathbf{r}),
\end{equation}
where $T$ is a normalization factor, is
commonly referred to as the trilobite electronic state due to its the uncanny resemblance to the structure of a trilobite fossil (as seen in Fig.~\ref{fig1}(b)). For large $n$-values, the trilobite PEC can be divided into an anharmonic well overlayed with an oscillatory potential, i.e. $V_t(R) = V_{anh}(R) + V_{osc}(R)$.  The overall anharmonic envelope is very well represented by the approximation \cite{Greene2000,Omont_1977}
\begin{equation}    \label{Eq:V_anh}
    V_{anh} (R) = \frac{a_{s}^T(k)}{\pi n^3} \sqrt{ \frac{2}{R} -\frac{1}{n^2} - \frac{(l_{min} + 1/2)^2}{R^2}} .
\end{equation}
At large $R$ values beyond the final well in the PEC, $V_{anh}$ is no longer an accurate envelope of the trilobite potential.  However, we will limit our work here to ranges in which it is.
In the following work, we will focus on the wavepacket dynamics of the URLM in the case of trilobite molecules.

Inclusion of important corrections due to $p$-wave shape resonances, fine structure of the Rydberg atom, hyper-fine structure of the ground-state atom, and nonadiabatic interaction effects will undoubtedly produce quantitative changes of the electronic structure and related PEC of the trilobite molecule. However, the interesting dynamical phenomena predicted with the relatively basic approximations used in this work are expected to persist qualitatively in a more accurate treatment. It is also worth mentioning that a heterogenous ULRM \cite{killian2020,Deiglmayer2021}, consisting of an $^{87}$Rb Rydberg atom and a $^{84}$Sr ground-state atom neither exhibits a hyper-fine structure, nor a p-wave shape resonances \cite{killian2015,Camargo_2016}. Such a Rb-Sr molecule might be an ideal case to study the dynamical properties of pure trilobite molecules without excess complexities, and can prove useful in the empirical reproduction of phenomena discussed in this work.

\begin{figure*}
        \centering
        \includegraphics[width=0.8\linewidth]{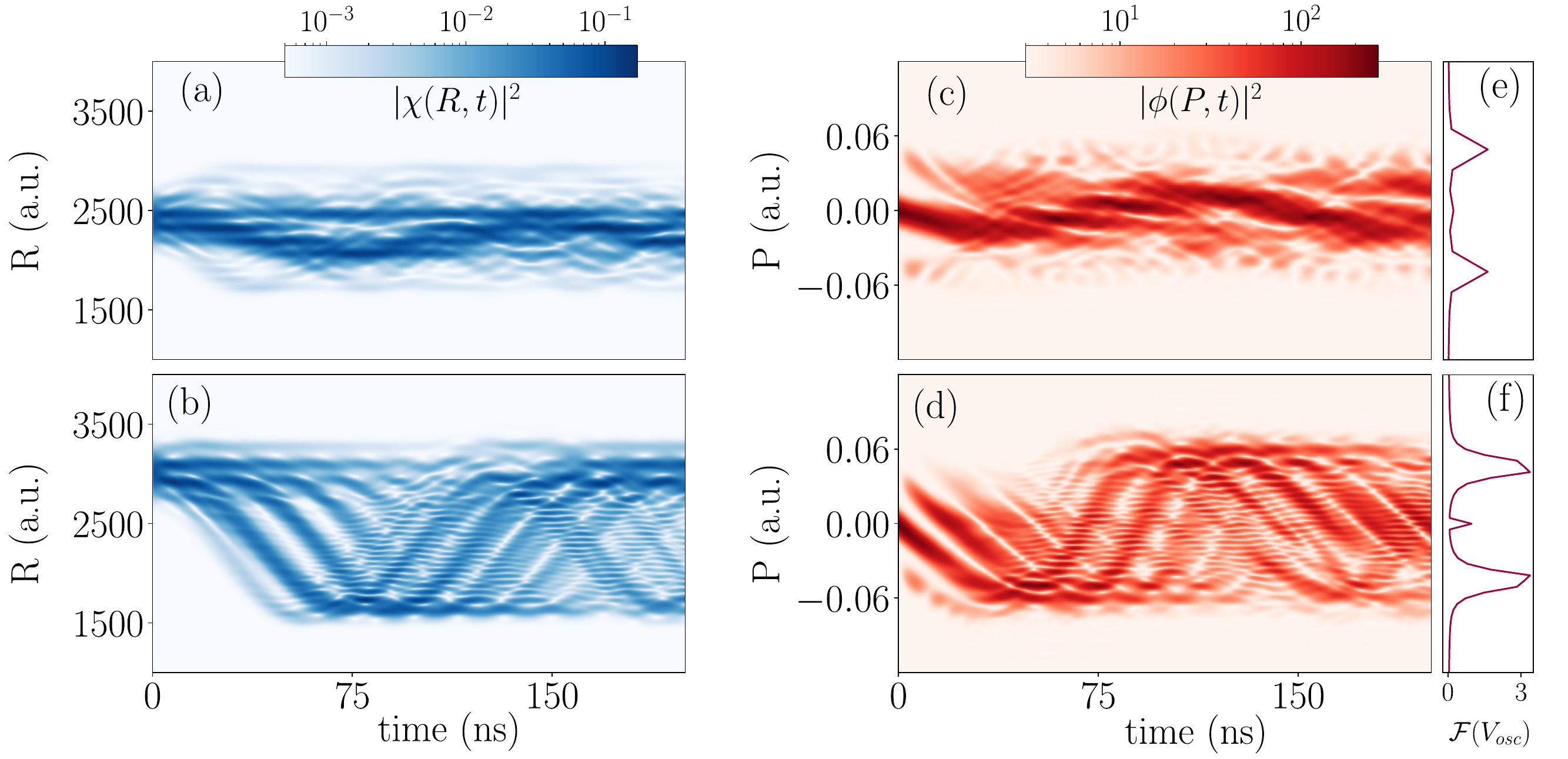}
        \caption{Onset of wavepacket diffraction. Probability density plots $|\chi (R,t)|^2$ and $|\phi (P,t)|^2$ of the wavepacket scattering off an $n$=55 trilobite potential, in the position (a-b) and momentum (c-d) representation. The initial state is a Gaussian wavepacket of characteristic width 100 a.u., centered around $R_0$=2400 (a) and $R_0$= 3000 (b) a.u.~with the corresponding momentum representation presented in (c) and (d) respectively. Panels (e-f) illustrate the corresponding Fourier transform of $V_{osc} (R)$ probed by each of the wavepackets.}

        \label{fig3}
\end{figure*}

\subsection{Vibrational dynamics}    \label{Sec22}

To explore the vibrational dynamics of a ULRM in the trilobite state, we solve the corresponding time-dependent Schrödinger equation given at the level of the Born-Oppenheimer approximation by:
\begin{equation}    \label{Eq:Schroedinger}
    i \hbar \frac{\partial}{\partial t} \chi(R,t) = H_{n} \chi(R,t),
\end{equation}
where $\chi(R,t)$ is the time dependent nuclear wavefunction representing the radial dynamics of the ULRM along the internuclear axis.  The nuclear Hamiltonian, $H_n$ is given in atomic units as
\begin{equation}     \label{Eq:H_n}
    H_{n} =  -\frac{1}{2\mu} \nabla_R^2  +  V_t(R),
\end{equation}
where the first term is the kinetic energy of the relative motion of the nuclei (with reduced mass $\mu$) and $V_t$ is the trilobite PEC from Eq.~\eqref{Eq:V_t}. We propagate the initial wavepacket $\chi(R,t=0)$ in time via the standard eigenstate expansion, i.e. 
\begin{equation}    \label{Eq:Chi}
    \chi(R,t)=\sum_k a_k e^{-iE_kt}\chi_k (R),
\end{equation}
where $\chi_k(R)$ are the stationary eigenstates of $H_n$ with eigenenergies $E_k$, i.e. $H_n \chi_k(R) = E_k \chi_k(R)$, and the expansion amplitudes are given by the standard inner-product $a_k=\left<\chi_k(R)|\chi(R,t=0)\right>$.

Our wavepacket is initialized as a Gaussian centered at position $R_0$ with width $\sigma$ with zero initial momentum, i.e. $\chi(R,t=0)\propto \exp\left[  -\left(R-R_0\right)^2/(4\sigma^2) \right]$, and the time dependent probability density function $|\chi (R,t)|^2$ is determined. An initially displaced wavepacket in the trilobite potential, will oscillate back and forth, returning to its initial position accounting for the anharmonic character of the potential.  During this process, it will scatter off the oscillatory portion of the potential, which acts as a diffraction grating, facilitating the back-scattering (and forward- scattering) of a time-evolving wavepacket. Via the Fourier transform, we also analyze the momentum space wavefunction, $\phi(P,t) = \mathcal{F}\{\chi (R,t)\}$, and the oscillatory potential $\mathcal{F}\{V_{osc} (R)\}$ to study and explain such scattering effects in momentum space. In the following we numerically investigate the vibrational dynamics of ULRM in the position and momentum space for different initial state configurations and different principle quantum numbers, using the framework discussed above. The stationary eigenstates of $H_n$ are calculated utilizing a tenth order finite difference method \cite{Groenenboom_1990}, which are then used to simulate the time-evolving wavepacket (Eq.~\ref{Eq:Chi}). Owing to the simplicity of the system, we achieve converged, accurate results with relatively little computational effort. Hence the results have been calculated with step sizes of $\Delta_R \sim 2 $ a.u. and $\Delta_t \sim 0.05 $ ns.

\section{Results and Discussion}   \label{Sec3}
 
In this section, we analyze the vibrational dynamics of an oscillating vibrational wavepacket in the trilobite molecule. We discuss the potential energy structure of trilobite molecules for multiple $n$-values, highlighting their oscillatory structure as a salient feature. We then proceed to illustrate the probability density plots of a time propagating Gaussian wavepacket as it scatters off a $n$=55 trilobite potential. The unique dynamical effects that arise as a function of the width and position of the initial wavepacket is demonstrated and explained utilizing the Fourier tranform of the wavepackets and the oscillatory features of the trilobite potential. Finally,  prototypical experimental schemes that might be used to observe such dynamical effects are also discussed.

\subsection{Potential energy landscape }   \label{Sec31}

The potential energy curves for the principal quantum numbers ranging from $n$=55 to $n$=69 are illustrated in Fig.~\ref{fig2}(a). The $R$ range is chosen to highlight the oscillatory structure exhibited by $V_t$. Solid curves represent the trilobite PEC (color-coded with respect to $n$), while dashed lines in the back-ground represent the approximate anharmonic potential $V_{anh}$ according to Eq.~\eqref{Eq:V_anh}. Subtracting $V_{anh}$ from the total trilobite potential allows us to isolate the remaining oscillatory part of the trilobite PEC, $V_{osc}$ shown in Fig.~\ref{fig2}(b). The potential energy curves exhibit an increasingly shallow envelope and amplitude of oscillation for higher values of $n$. We also observe a  slight increase in the frequency of oscillation for a corresponding increase in $n$.

Additional insight can be gained by analyzing the oscillatory potential in momentum space through the Fourier transform of $V_{osc} (R)$, $\mathcal{F}\{V_{osc} (R)\}$, the results of which are shown in Fig.~\ref{fig2}(c). In momentum space representation, the potentials exhibit a clear side-band structure that corresponds to its undulating nature in $R$. For example, the $n$=55 trilobite state exhibits oscillations in $R$ whose wavelength is approximately 100 a.u., though it does vary increasing up to $\sim 230$ a.u in the $R$-range considered here. This results in a Fourier spectrum that is strongly peaked at $P$ = 0.03 a.u., where $P$ is the radial momentum in the Fourier space. As the oscillations of the PEC deviates from a strict periodic behaviour in position space, we observe a broadening of the side-band structures with an overlay of multiple peaks in Fourier space. Note that the Fourier transform is performed using a truncated range of $R$-values ranging from $R$=1300 a.u. to 4500 a.u. This corresponds to the range of internuclear separations probed by the oscillating wave-packet. The extremes of the position of the peak of the wavepacket are determined by the classical turning points of the system in the trilobite potential when the wavepacket is started from rest displaced to one side. The usage of such a limited grid in position space restricts the accuracy of the Fourier analysis and invariably contributes to further broadening of the side-bands and the exact positions of the multiple peaks. Finally, the $n$-dependence of the amplitude of oscillations of the PEC as discussed before, is also reflected in the $n$-dependence of the amplitude and position of the side-bands in Fig.~\ref{fig2}(c).

\subsection{Vibrational dynamic effects}   \label{Sec32}

\begin{figure}
        \centering
        \includegraphics[width=\linewidth]{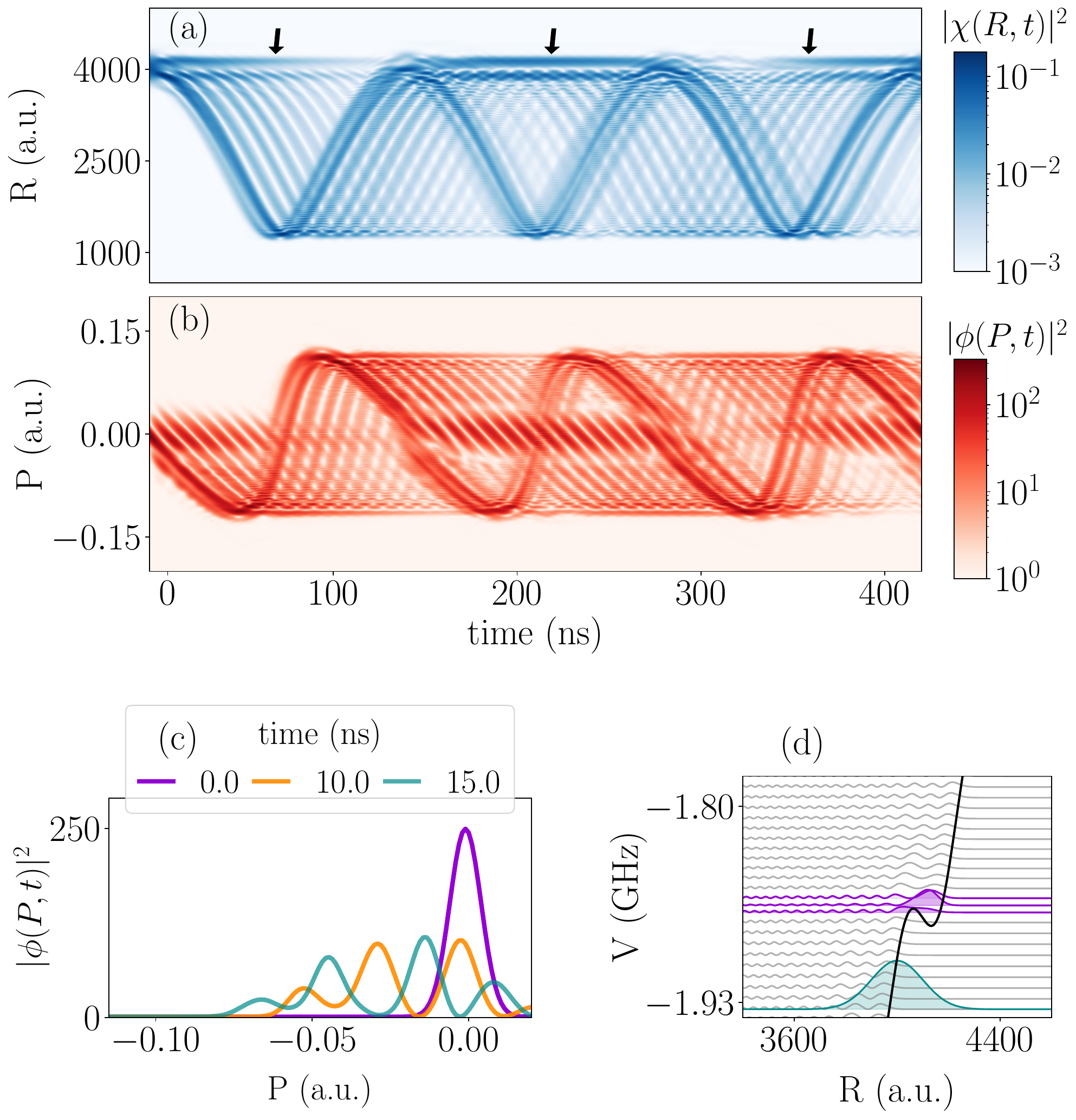}
        \caption{ Probability densities $|\chi (R,t)|^2$ and $|\phi (P,t)|^2$ of the wavepacket scattering off an $n$=55 trilobite potential, in the position (a) and momentum (b) representation. The initial state is a Gaussian wavepacket of width 100 a.u., centered around $R_0$= 4000 a.u. The arrows mark the stickiness of the wavepacket discussed in the text. (c) Time-slices of $|\phi (P,t)|^2$, illustrating the back-scattering undergone by the propagating wave-packet. (d) The probability densities of the molecular eigenstates, with the localized states responsible for the initial ``stickiness'' highlighted (purple) near the initial Gaussian (cyan).}

        \label{fig4}
\end{figure}
\textit{Onset of wavepacket diffraction -} We begin by performing a series of dynamical simulations of the ULRM’s nuclear wavepacket initialized in a Gaussian state centered around different values of $R$, in an $n=55$ trilobite PEC. The wavepackets are constructed such that the width of the corresponding probability density exhibits a standard deviation of 100 a.u., which is comparable in size to the wavelength of the oscillations in the trilobite potential. Figure \ref{fig3} illustrates the probability density functions of the time evolving wavepacket, in the position space (Fig.~\ref{fig3}(a-b)) and momentum space (Fig.~\ref{fig3}(c-d)).
 The position of the initial state determines the total energy of the time evolving wavepacket, thereby putting an upper bound on the magnitude of radial momentum probed by the nuclei during its dynamics. A Gaussian state centered around $R_0=2400$ a.u (Fig.~\ref{fig3}(a,c)), i.e.~relatively close to the equilibrium internuclear distance ($R_{eq} \sim$ 2190 a.u), performs low energy oscillations around $R_{eq}$ with complex multi-well interference effects. Interestingly, a further increase in the total energy of the wavepacket, achieved by having its initial position extended to $R_0=3000$ a.u (Fig.~\ref{fig3}(b,d)), exhibits the onset of a clear multiple peak structure that resembles a time propagating diffraction pattern. The emergence of these unique interference patterns is explained by focusing on the momentum picture. Oscillations that achieve lower momentum, are not able to probe the clear side-band structure of the trilobite potential exhibited by its oscillatory features (as seen in Fig.~\ref{fig3}(e)). Hence they only display very faint forward and back-scattering peaks that are quickly dispersed in the course of time. However, oscillations as portrayed in Fig.~\ref{fig3}(d), quite clearly achieve higher momentum beyond the side-band structures (as seen in Fig.~\ref{fig3}(f)), thereby showing clear signs of forward and back-scattering in the momentum picture. Such an explanation treats the oscillatory structure of the trilobite potential as a diffraction grating in momentum space. 

 \textit{Internal diffraction dynamics -} To emphasize the phenomena discussed above, we analyze the vibrational dynamics of a ULRM with an initial Gaussian wavepacket centered at $R_0=4000$ a.u. again with a width of 100 a.u. As discussed earlier, increasing $R_0$ farther away from $R_{eq}$ further increases the momentum probed by the wavepacket allowing it to clearly probe the momentum side-band structure of the potential as it scatters off the PEC. Hence, following the trend from the previous simulations (with $R_0=2400,3000$ a.u.), we see that initializing the wavepacket with $R_0=4000$ a.u. facilitates repeated coherent back-scattering as the wavepacket traverses through the PEC. The resulting time-dependent probability density is shown in Fig.~\ref{fig4}(a). In this case the system samples enough of the oscillatory potential to constantly undergo forward and back-scattering of the oscillatory part of the trilobite potential produces a train of roughly evenly spaced side peaks. The origin of this behavior becomes apparent in the momentum space probability density shown in Fig.~\ref{fig4}(b).  Here, the initial momentum space wavepacket scattering off the oscillatory potential producing adjacent scattered peaks corresponding to the momentum of the peak of the Fourier transform. Figure \ref{fig4}(c) shows time slices of the momentum probability density function at different times showing peaks separated in momentum by around $\sim$ 0.03 a.u, approximately the peak of the side-bands seen in the momentum representation of the PEC in Fig.~\ref{fig2}(c). The resulting fragmented wave-train structure in momentum space exhibits a clear and coherent multi-peak structure, reminiscent of that observed in waves scattering from a diffraction grating. Apart from the diffraction effects, we also notice that the probability density plots in Fig.~\ref{fig4}(a) (marked by the arrows), exhibit sustained population near the initial state of the wavepacket which we denote as ``stickiness". This is a result of a large overlap of the initial wavepacket with eigenstates that are localized around a well near the initial position (see Fig.~\ref{fig4}(d)). Figure \ref{fig4}(d) shows the vibrational probability densities of eigenstates that participate in the wavepacket dynamics as well the initial wavepacket (cyan) for reference. These localized states (purple) have a large overlap with the initial wavepacket and exhibit a vibrational splitting around $\sim 46$ MHz, as opposed to the delocalised states (grey) exhibiting a vibrational splitting of $\sim 67$ MHz. Part of the wave-packet evolves with the time-scales associated with the localized states ($\sim 210$ ns), which is noticeably larger than the time-scale of the anharmonic oscillation ($\sim 150$ ns), thereby exhibiting said stickiness. Due to the limited amount of eigenstates involved, this particular phenomena continues to occur for even larger values of $t$ ($\geq$ 400 ns), retaining its spatial localisation periodically, similar to a quantum revival. However, the propagating diffraction pattern is continuously scattered off of a band of varying frequencies diminishing the chance of a complete quantum revival in finite time.

\begin{figure}
        \centering
        \includegraphics[width=\linewidth]{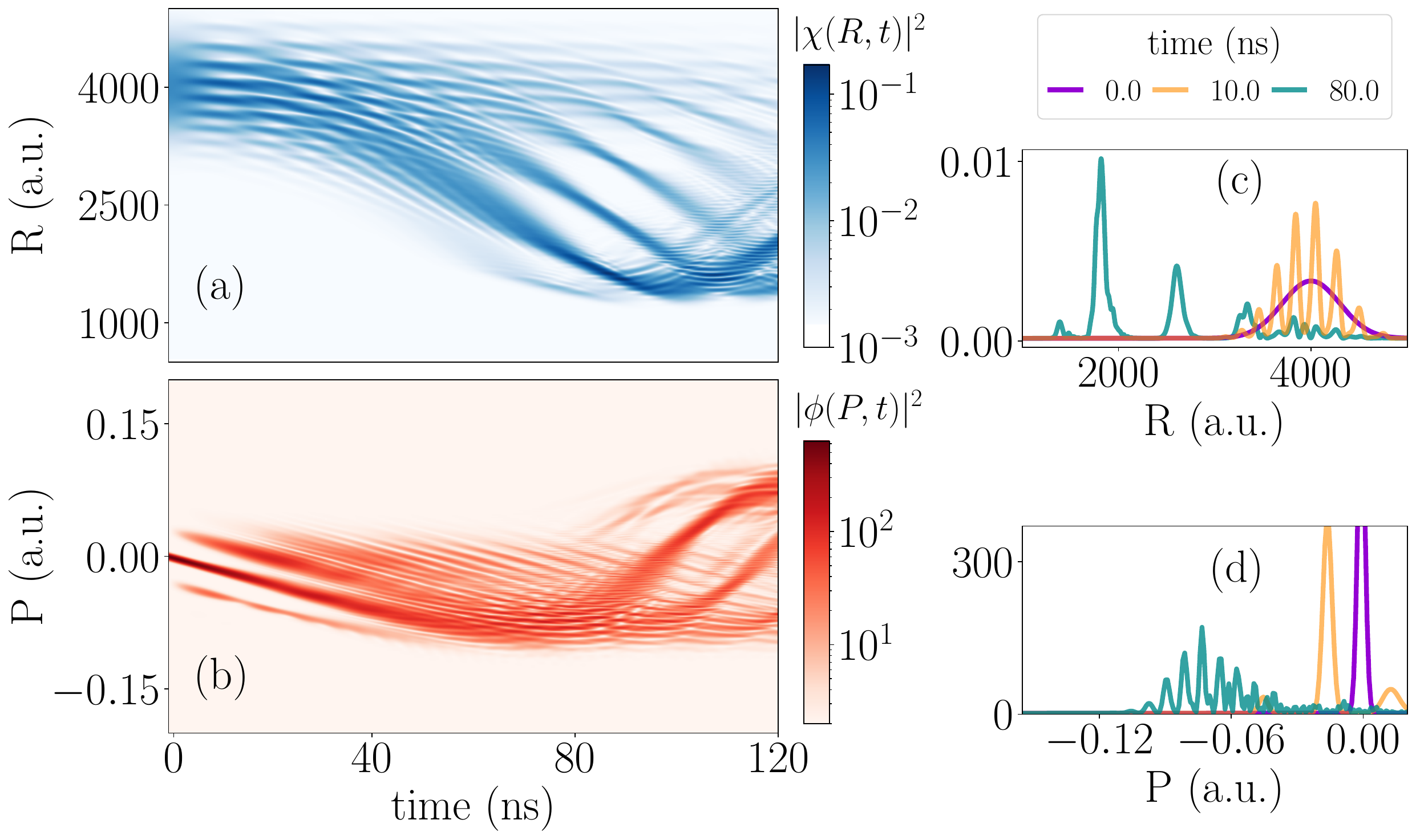}
        \caption{Interference effects in the scattering of a wide wavepacket. Probability densities $|\chi (R,t)|^2$ and $|\phi (P,t)|^2$ of the wavepacket scattering off an $n$=60 trilobite potential, are shown in the position (a) and momentum (b) representation. The initial state is a Gaussian wavepacket of width 300 a.u., centered around $R_0$=4000 a.u. Time-slices  of $|\chi (R,t)|^2$ and $|\phi (P,t)|^2$ illustrating key stages of evolution are shown in the panels (c) and (d), respectively.}

        \label{fig5}
\end{figure}

\textit{Wave train propagation -} The aforementioned time propagating interference effects are dependent on the width of the nuclear wavepacket. Above, we used an initial state whose width was of comparable size to the wavelength of the osillations of the PECs. A spatially very narrow wavepacket does not experiences the interference effects seen above, and primarily perform anharmonic oscillations, with dispersion enabled by the undulating structure of the PEC. On the other hand, a wavepacket that is spatially considerably wider than the wavelength of the oscillations in $V_t$ will show dynamical phenomena distinct from those discussed above. Figure \ref{fig5} illustrates the dynamics of an initial Gaussian wavepacket centered at $R_0=4000$ a.u. propagating in the $n=60$ trilobite curve. The $n=60$ trilobite features an oscillatory potential with a comparatively smaller wavelength than that of $n=55$. The initial state is constructed with a width of 300 a.u.,  several times larger than the well separation distance of $V_t$. Hence, at $t=0$ the wavepacket spans multiple wells of the trilobite potential. During the initial time dynamics each well of the PEC acts as a source creating a multipeaked fragmented wavepacket with each peak localized at a different well, as portrayed in the $t =8$ ns time slice shown in Fig.~\ref{fig5}(c). Later in the course of the dynamics, these individual wavepackets emanating from different wells interfere with each other forming very narrow localized pulses ($t =80$ ns, Fig.~\ref{fig5}(c)) that are isolated from each other. The width of these narrow peaks do not conform to the size of the initial wavepacket, and are dependent on the scattering potential in which the wavepacket propagates. Their formation can be explained by considering the momentum space picture shown in Fig.~\ref{fig5}(b). Here we see that an extremely narrow initial wavepacket in momentum space undergoes constant back-scattering to form an approximately equidistant train of sharp peaks in $P$-space ($t =80$ ns, Fig.~\ref{fig5}(d)), loosely reminiscent of a Dirac comb \cite{Cordoba1989,diddams1999broadband}. The Fourier transform of such a structure exhibits a pulse-train behaviour such as that observed above. As a final note, for higher $n$-values, both the amplitude of oscillations $V_{osc}$ and the general anharmonic envelope $V_{anh}$ of the trilobite PEC becomes considerably more shallow. Hence, the momentum range probed by the wavepacket and the magnitude of back-scattering it experiences are reduced, thereby resulting in a diminished diffraction pattern.

\subsection{Limitations of the theoretical framework} \label{Sec33}

We have ignored the effects of $p$-wave shape resonances and spin-structure in our analysis.  Inclusion of the $p$-wave scattering term would result in a more complex PEC structure that features avoided crossings between trilobite and butterfly curves near $R_{eq}$. These avoided crossings would substantially alter the long-time dynamics of the wavepacket as it traverses through the full potential energy landscape. This might lead to the introduction of vibronic decay, surface hopping, diabatic stabilization and several other nonadiabatic effects. Recent works have explored the spectral features of such systems \cite{Hummel_2022,Srikumar2023,durst2024} and have shown that nonadiabatic effects are of increasing importance at higher $n$-values (like $n\geq$ 55) \cite{Raithel2022}. However, our present work is a proof of methodology meant to bring attention and interest to the study of wavepacket dynamics of trilobite molecules. Furthermore, while the $p$-wave shape resonances will substantially alter the long time-scale dynamics, the initial time dynamics should still qualitatively exhibit the diffractive effects discussed. Also, one can mitigate the complexity introduced by spin-structure to some extent by careful preparation and state selection of the hyperfine state of the ground-state atom. Moreover, the hyperfine splitting of the PECs introduced by ground-state $^{87}$Rb remains around $\sim$ 6.8 GHz while the depth of the trilobite potentials scale as $n^{-3}$ with increasing $n$. At high enough $n$-values, for example $n \geq$ 55, the rather large energetic separation between the different hyperfine states can be utilised to select and probe a single PEC. Of course, it is important to note that these compilcations can be eliminated by selecting a different ground-state perturbing atom, such as Sr which has no low energy, p-wave electron-atom resonances and lacks any hyperfine structure.

\subsection{Prototypical experimental setup} \label{Sec34}

The numerical simulations considered have been focusing on the time evolution of Gaussian wavepackets. Experimentally one can utilize an optical tweezer setup \cite{Schlosser2001,Kaufman2012} which confines two ground-state Rb atoms in separate harmonic potentials. The width and position of the initial Gaussian wavepacket can be tuned through the depth of the optical traps and the separation between the two traps, respectively. By employing a multi-photon excitation of an individual atom to the trilobite state while simultaneously turning off the confinement potential creates the desired initial state of the ULRM. Such a scheme, although straightforward, is difficult to accomplish in practice utilizing standard optical tweezers. Optical setups with finite laser waists cannot probe the nanometer scale confinement required to prepare motional states whose dimensions are of the order of $\sim$ 100 a.u. Advancements in cold atom trapping employing the integration of plasmonic tweezers that facilitate nanoscale confinement may be one future option to circumvent this problem \cite{Mildner_2018,Zhang2021,Crozier2019}.

\begin{figure}
   \includegraphics[width=0.46\textwidth]{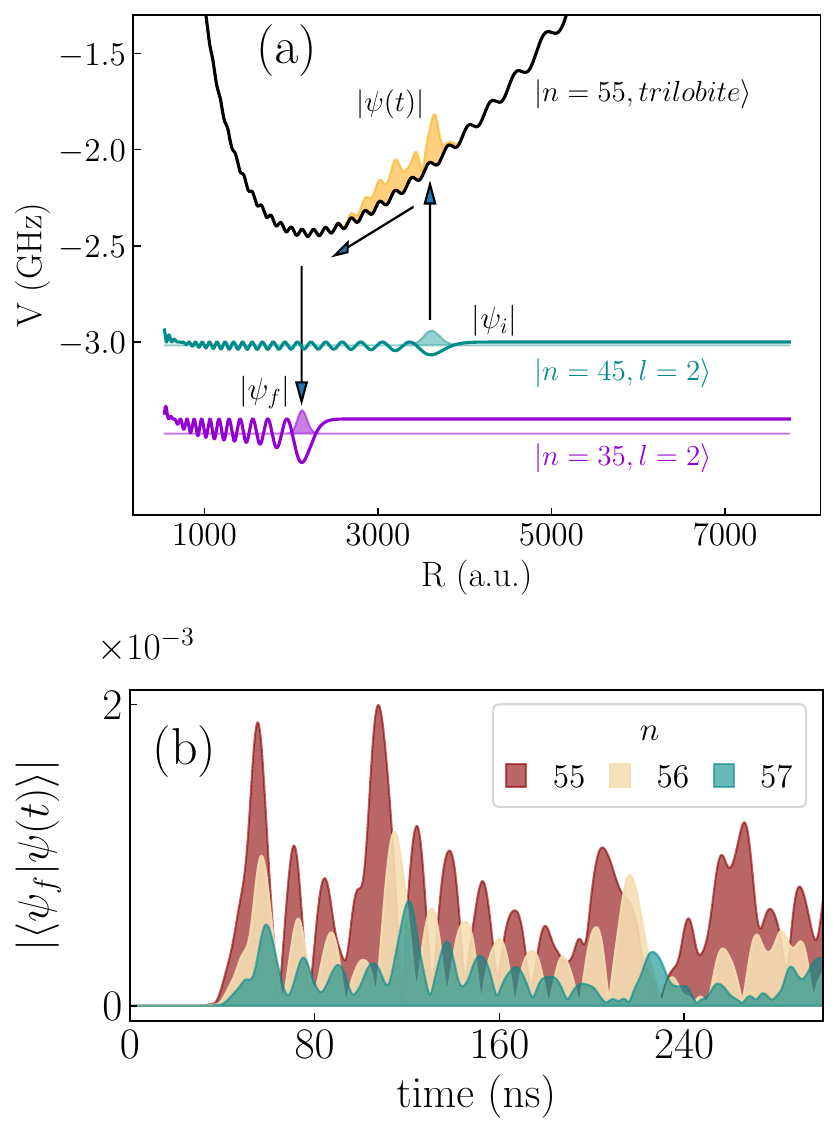}
    \caption{Pictorial scheme of the suggested pump-probe experiment. (a) The vibrational ground state of the $\ket{n=45,l=2}$ ULRM is pumped to the $n$=55 trilobite potential and subsequently propagates therein. Following a time lag the wave packet is projected onto the desired $\ket{n=35,l=2}$ probe state. (b) Observable time resolved correlation function calculated for multiple $n$-values using the given setup. For each $\ket{n,trilobite}$ PEC, the pump and probe states are selected to be $\ket{n-10,l=2}$ and $\ket{n-20,l=2}$ respectively, where $n\in \{ 55,56,57 \}$.}
    \label{fig6}
\end{figure}

With this in mind, let us propose an alternate experimental scheme that utilizes a time-resolved pump probe methodology to monitor dynamical effects in real time. The basic idea is to utilize lower-lying low-$l$ ULRM to create the initial state and as a probe to monitor the system. Figure \ref{fig6}(a) illustrates a prototypical example of the proposed scheme. The initial state preparation is accomplished by a two photon transition to the Gaussian-like vibrational ground-state of the $\ket{n=45,l=2}$ ULRM state. The vibrational ground-states for different principle quantum numbers feature a convenient range of mean positions and widths, providing the necessary flexibility needed to observe the dynamical effects we seek. They can also be conveniently used as a probe-state localized near a range of desired $R$-values. Similar pump-probe schemes utilising low-$l$ Rydberg states were also utilized in the study of ion-atom scattering recently \cite{Pfau2018,Duspayev_2021}. As an example, 
the ULRM is excited from the $\ket{n=45,l=2}$ state to the $n$=55 trilobite electronic state using a microwave pump pulse, utilising $f$-state admixture of the high-$l$ molecule \cite{Althoen2023}.  The microwave excitation is fast enough to consider it projective, preserving the vibrational wavepacket as the initial state that porpagates in the trilobite potential. A time-delayed probe-pulse then enables us to project the time evolving wavepacket onto a $\ket{n=35,l=2}$ probe-state and obtain the time resolved correlation function. Figure \ref{fig6}(b)shows the predicted result of this proposed experiment for three different $n$-values. In each case the $\ket{n-10,l=2}$ pump-state is excited to the $\ket{n,trilobite}$, with $\ket{n-20,l=2}$ being the probe-state. The state selection is done to facilitate comparable propagation times and correlation functions, as seen in Fig.~\ref{fig6}(b). The train of evenly spaced diffraction peaks produced by scattering of the wavepacket from the trilobite potential is observed as a series of peaks in the time resolved correlation function. Note that the amplitudes of the correlation function, which carries the phase-information of the diffraction pattern, decreases with increasing $n$. Although not a systematic study of the $n$-dependence of the diffraction effect, the decrease in the amplitude of correlation function is consistent with our analysis that higher $n$-values exhibit reduced back-scattering and consequently display diminished diffraction effect. Rather conveniently, the temporal resolution achieved here is only limited by the time delay between the pump and probe pulses and their respective widths, both being finely tunable in the nanosecond regime. In comparison, direct spatio-temporal imaging as achieved in state-of-the-art ion microscopes \cite{Zuber2022,Pfau2021} limits the resolution to the order of $\sim 100$ nm and $\sim 10$ ns, rendering the desired phenomenon challenging to observe. Moreover, recent advances in the construction of long-range molecular potentials with tunable electronic character \cite{Niederprum2023}, allow the possibility of engineering initial and probe states that can be optically excited to a trilobite state of choice. Note that although the theoretical description of a pump-probe scheme is rather straightforward (and abundant in conventional molecular physics), a corresponding experimental realization of the scheme in ultracold systems is not without due challenges \cite{Pfau2018,Duspayev_2021}. But, within the scope of this work the pump-probe framework allows for the initial state preparation, time propagation, and observation of quantum interference and diffraction effects all encompassed within a single molecular system.

\section{Conclusion and Outlook}    \label{Sec4}

We have investigated intriguing wavepacket interference effects that unfold in the nuclear dynamics of a single diatomic trilobite ULRM.  The internuclear potential landscape mediated by the Rydberg electron exhibits rapid oscillatory features similar to those of a diffraction grating. We observe that trilobite molecules with nuclear wavepacket widths comparable to that of the characteristic wavelength of the PEC, propagating with high enough initial energy will undergo back-scattering as a consequence of its own electronic structure. An elementary Fourier analysis of the propagating wavepacket and the potential energy landscape explains the observed phenomena, with semi-quantitative detail. The self diffraction phenomena exhibited by these molecules can be readily observed by an appropriate time resolved pump-probe scheme. Conveniently, the necessary initial state-preparation is accessible via the vibrational ground-state of e.g.~lower-lying $l=2$ ULRM states. Hence we suggest that the ULRM can be functionally used as a single molecular laboratory for observing cold-atom dynamics on exaggerated time and length-scales. The behaviors we predict within the simplified description used here should persist qualitatively in the full system and should be readily observable experimentally.

Our work provides new windows of opportunity for the study of dynamical behaviour of ultralong-range Rydberg molecules. An immediate extension of our work, featuring the inclusion of $p$-wave shape resonances of Rb and the consequent introduction of vibronic couplings in the system, offers a new pathway to study multichannel dynamical effects in cold-atom systems. The high-$l$ nature of the trilobite and butterfly configurations exhibiting large dipole transition elements might give rise to a plethora of dynamical effects not usually observed in standard atomic or molecular systems. Furthermore, the massive dipole moment of the trilobite state (and the butterfly state) \cite{Chibisov_2002,Booth_2015,Hamilton_2002} can be skillfully exploited by the introduction of an external electric field. This opens the possibility of new explorations in wavepacket dynamics near Conical Intersections \cite{Hummel2021}, dynamics near pendular states \cite{Niederprum2016}, and the coherent control of ULRM to achieve rovibrational state engineering. 

\begin{acknowledgments}
We are grateful for the financial support granted by the German Research Foundation (DFG)  through the priority program ``Giant Interactions in Rydberg Systems" [DFG SPP 1929 GiRyd project SCHM 885/30-2]. 
\end{acknowledgments}

\bibliographystyle{apsrev4-2}

\end{document}